\documentstyle[12pt]{article}
\def\ba{\begin{eqnarray}}
\def\ea{\end{eqnarray}}
\def\lb{\label}
\def\be{\begin{equation}}
\def\ee{\end{equation}}

\begin{document}
\begin{titlepage}
\title{The Differential Calculus on Quantum Linear Groups.}
\author{  L.D.Faddeev
\thanks{Supported by the Russian Academy of Sciences and Academy of Finland.}
\thanks{e-mail address: faddeev@imi.spb.su}\\
\it
St. Petersburg Branch of the Steklov Mathematical Institute, \\
\it Fontanka 27, St. Petersburg 191011, Russia \\
\rm and \\
P.N.Pyatov
\thanks{Supported in part by the Russian Foundation of Fundamental Research
(grant 93-02-3827).}
\thanks{e-mail address: pyatov@thsun1.jinr.dubna.su}
\\
\it Bogolubov Theoretical Laboratory, \\
\it Joint Institute for Nuclear Research, \\
\it 141980 Dubna, Moscow Region, Russia}
\date{}
\maketitle

\begin{abstract}
The non-commutative differential calculus on
the quantum groups $SL_q(N)$ is constructed.
The quantum external algebra proposed contains
the same number of generators as in the classical case.
The exterior derivative defined in the constructive way
obeys the modified version of the Leibnitz rules.
\end{abstract}

\end{titlepage}
\newpage

\section{Introduction}
\setcounter{equation}0

Recent interest in constructing the differential calculi on
the quantum groups stems from the Woronowicz's
pioneering work \cite{Wor2}. There he has formulated
the general algebraic framework for dealing with the problem.
In the subsequent investigations the emphasise was made on
two main directions. First, the experience of dealing with
such algebras was accumulated while considering the simplest
lower dimensional examples (see e.g. \cite{Wor1,PW,SVZ}).
It was soon recognized that the true quantum group differential
calculus should be bicovariant, and that this condition is very
restrictive. Indeed, the only use of this condition allows
one to obtain the unique external algebra construction for
the $SL_q(2)$ Cartan $1$-forms \cite{IP1}. Next, a very close
connection was established with the theory of quadratic quantum
algebras (quantum spaces) \cite{Man0,FRT,WesZum}. It was then realized that
the condition of unique ordering of higher order monomials (the so-called
diamond condition) is very important \cite{Man2,Malt}, and that really
it is only to be checked for the cubic monomials \cite{Man1}.

Another direction of investigations was finding out the
adequate technique for dealing with quantum differential algebras.
Here the close connections between the quantum differential calculi
and the $R$-matrix formulation for quantum groups and algebras
\cite{FRT} were soon established \cite{Jurco,Faddeev,Zum1}
(for further  considerations see \cite{AC}). It turns out that
the $R$-matrix technique is highly appropriate in treating the
problems arised.

The next stage of investigations was to combine both the lines of
research to obtain the concrete differential algebra constructions
for known series of quantum groups. Here the substantial progress was
achieved for the $GL_q(N)$ case. Namely, in the series of papers
\cite{Malt,Sud,Schir,SWZ,Tzy} the pair of the nice-looking differential
algebras on $GL_q(N)$ was constructed. But the situation with
the $q$-deformed series of the simple Lie groups appears to be
much more complex. The natural way of obtaining the $SL_q(N)$,
$SO_q(N)$ and $SP_q(N)$ differential calculi by performing reduction
>from the $GL_q(N)$ calculi failed in the quantum case, because
one cannot consistently reduce the number of the generating elements
in the $GL_q(N)$ differential algebras constructed (see discussion
in \cite{CSWW,Zum2}).
In principle one can treat these nonreduced (or partially reduced)
differential calculi as a quantizations of the nonstandard
classical calculi on the special groups ( see \cite{Mul}),
but the problem of finding out the deformations of the ordinary
calculi still remained open.
It is rather natural in this situation
to revise once again the basic postulates involved into constructive
scheme. The only postulate that seems too restrictive
is the classical Leibnitz
rule for the exterior derivative \cite{Faddeev}
$$
d(f\cdot g) = df \cdot g + (-1)^{|f|} f \cdot dg \; .
$$
Indeed, let us remind that the basic vector fields after quantization
correspond to the finite shifts rather than to infinitesimal
differentiations. The natural Leibnitz rule for them is
multiplicative instead of being additive. Correspondingly
the Leibnitz rule for differential must take into account
this shift property of vector fields.

In this paper we propose the construction of the differential
algebra with properly modified Leibnitz rule. We consider
the case most close to $GL_q(N)$ -- the $SL_q(N)$ differential algebra.
Here only one Cartan $1$-form and one basic vector field should be reduced.
The reduction scheme for vector fields was already developed
in \cite{SWZ}. We propose the reduction scheme for Cartan $1$-forms.
%The price is the modification of Leibnitz rules, but it is very natural
%to do so since after quantization the structure of the quantum group
%manifold is discretized, and we have no differentiation, but finite
%shifts on it.
We do not discuss here the involution, leading to the unitary
reduction of our system. As was shown in  \cite{AF2}
this can be done for $q$ on the circle ($|q| = 1$) for the algebra
of vector fields and functions on the quantum group.
We believe, that the
involution , found in \cite{AF2} can be continued on
the differential forms as well.

The paper is organized as follows. In Section 1 we fix the notations
of the $R$-matrix technique and formulate the basic posulates of our
construction. We believe that it was the consistent use of the
$R$-matrix technique which allowed us to make the construction
through. So it not only led to the simplifying of calculations, but
played an important heuristic role.
In Section 2 we present the external algebra on $SL_q(N)$. This
algebra is also supplied with the action of the basic vector
fields (or Lie derivatives). We refer to this extended algebra as
the differential algebra on $SL_q(N)$. Section 3 is devoted
to construction of the exterior derivative operator $d$.
Note that the scheme proposed can be equally applied to $GL_q(N)$.
In this way one can recover a wide variety of the differential
algebras on $GL_q(N)$. It seems to us that such a nonuniqueness
is due to the nonsemisimplicity of $GL_q(N)$.

\section{The basic principles and notation}
\setcounter{equation}0

$\;$The starting point for our consideration is the$\;$ Hopf algebras$\;$
$Fun(GL_q(N))$$\;$ and $Fun(SL_q(N))$ \cite{FRT}. We present here
some facts and definitions for these algebras.

We choose the corresponding $R$-matrix \cite{Jimbo}
$R \in Mat_N({\bf C})^{\otimes 2}$ in the form
\be
\lb{1}
R = q \sum_{i} e_{ii} \otimes e_{ii} +
\sum_{i \neq j} e_{ji} \otimes e_{ij} +
\lambda \sum_{j<i} e_{jj} \otimes e_{ii} \; ,
\ee
where $i,j = 1, \ldots , N$ and $\lambda = q - {1 \over q}$.
In what follows  we will also use the shorthand notation $R$
for the matrix $R \otimes I \in Mat_N({\bf C})^{\otimes 3}$,
where $I \in Mat_N({\bf C})$ is the unit matrix. One can easily
distinguish, in the context of each formula, whether  $R$ belongs
to $Mat_N({\bf C})^{\otimes 2}$ or to $Mat_N({\bf C})^{\otimes 3}$.
The $R$-matrix (\ref{1}) satisfies the Yang-Baxter equation and the
Hecke condition, respectively,
\ba
\lb{YB}
R \, R' \, R &=& R' \, R \, R' \; , \\
\lb{H}
R^2 &=& {\bf I} + \lambda R \; .
\ea
Here $R' = I \otimes R$, and ${\bf I} = I \otimes I$. It is worthwhile
to establish the connection with other $R$-matrix conventions of
frequent use:
$$
\begin{array}{ll}
{\rm our} \;\; R \; & = \; {\hat R}_{12} \; = \; P_{12} R_{12} \; = \;
		     R^{+}_{12} P_{12} \; , \\
{\rm our} \;\; R^{-1} & = \; R^{-}_{12} P_{12} \; .
\end{array}
$$
Here $P \in Mat_N({\bf C})^{\otimes 2}$ is the permutation matrix and
all the notation ${\hat R}_{12}$, $R_{12}$, $R^{\pm}_{12}$ is given
in Ref. \cite{FRT}.

The unital associative algebra $\;Fun(GL_q(N))\;$ is generated by $\;N^2\;$
elements $T=(t_{ij})_{i,j=1}^{\;\;\;\;\;\;N}$.
The multiplication and
comultiplication in it are defined, respectively, by
\ba
\lb{RTT}
R \, T \, T' &=& T \, T' \, R  \; , \\
\lb{comult}
\triangle (t_{ij}) &=&  t_{ik} \otimes t_{kj} \; ,
\ea
where $T$ means $T \otimes I$ in (\ref{RTT}) and $T' = I \otimes T$.

The $q$-deformed Levi-Civita tensor  $\epsilon_{q}^{i_1 \ldots i_N}$
( $= \epsilon_{q}^{1 \ldots N}$ in brief notation) satisfies
the following characteristic relations:
\be
\lb{e}
\epsilon_{q}^{1 \ldots N} R_i = - {1 \over q} \epsilon_{q}^{1 \ldots N}
\; , \;\;\;\; 1 \leq i \leq N \; ,
\ee
$$
\epsilon_{q}^{i_1 \ldots i_N} \mid_{i_1 =1, \ldots , i_k =k, \ldots ,
i_N =N} \;= \; 1 \; .
$$
Here $R_i = I^{\otimes (i-1)} \otimes R \otimes I^{ \otimes (N-i-1)}$
(note: $R_1 = R$, $R_2 = R'$). The quantum determinant of $T$,
$det_q T$, defined through the relation
\be
\lb{det}
\epsilon_{q}^{1 \ldots N} T_1 T_2 \ldots T_N \; = \;
T_1 T_2 \ldots T_N \epsilon_{q}^{1 \ldots N} \; = \;
\epsilon_{q}^{1 \ldots N} \cdot det_q T \; ,
\ee
where $T_k = I^{\otimes (k-1)} \otimes T \otimes I^{\otimes (N-k)}$,
is the central element of the algebra $Fun(GL_q(N))$. This can be checked
by  the following formula
\be
\lb{eR}
\Psi^{N+1} \epsilon_{q}^{1 \ldots N} R_{N}^{\pm 1} \ldots R_{1}^{\pm 1} =
q^{\pm 1} \Psi^1 \epsilon_{q}^{2 \ldots N+1} \; ,
\ee
where $\Psi = (\psi^i)_{i=1}^{\;\;\;\,N} \in {\bf C}^N$ is an arbitrary
vector.
The Hopf algebra  $Fun(SL_q(N))$ is then obtained by adding
one more relation
\be
\lb{det=1}
det_q T = 1 \; .
\ee
to (\ref{RTT}). Finally, the antipodal mapping
$S(\cdot)$ on $Fun(GL_q(N))$ and
$Fun(SL_q(N))$ (for its explicit form see \cite{FRT})
satisfies the relations
\be
\lb{antipod}
S(T)T = TS(T) = I \; ,
\ee
therefore, in what follows we prefer using the notation $T^{-1}$ rather
than $S(T)$.
\bigskip

Now let us turn to the differential algebra of extensions of
$Fun(GL_q(N))$ and $Fun(SL_q(N))$. First, we should fix the basic
principles of our construction:
\medskip   \\
{\bf A.} {\it The bicovariance condition.} Following
\cite{Wor2} we require that a differential algebra should possess
the bicomodule structure with respect to the underlying
quantum group. By this condition we guarantee that the left and
right translations in a quantum group do not affect the structure
of its differential calculus. From this point it looks most
natural to use, say, right-invariant and left-adjoint vector
fields $\;L = (l_{ij})_{i,j=1}^{\;\;\;\;\;\;N}\;$ and Cartan 1~-~forms
$\;\Omega = (\omega_{ij})_{i,j=1}^{\;\;\;\;\;\;N}\;$ in addition to
$T$s  as the generating elements for differential algebra\footnote{
For the left-invariant and right-adjoint
generators all the constructions
proceed analogously.}.
The left anf right $Fun(GL_q(N))$-coactions in this case read:
\be
\lb{ad}
\delta_L(x_{ij}) = t_{ik}t_{lj}^{-1} \otimes x_{kl} \;\;\; ,
\;\;\; \delta_R(x_{ij}) = x_{ij} \otimes 1 \; ,
\ee
where by $X = (x_{ij})_{i,j=1}^{\;\;\;\;\;\;N}$ we understand
either $L$ or $\Omega$.

In case of the $SL_q(N)$ - differential algebra the number of
independent Cartan $1$-forms should be reduced by $1$.
This can only be achieved in a bicovariant manner with the use
of the $q$-deformed trace \cite{FRT,Reshet1} (see
also \cite{Zum1,SWZ,IM}). Here we define this operation and
present several useful formulae
\be
\lb{Tr}
Tr_q(X) = Tr({\cal D}X) \;\;\; , \;\;\;
{\cal D} = diag \{ q^{-N+1}, q^{-N+3}, \ldots ,  q^{N-1} \} \; .
\ee
The $Tr_q$-operation possesses the invariance property
\be
\lb{tr-inv}
Tr_q \delta_L (X) = 1 \otimes Tr_q X \; ,
\ee
and also
\ba
\nonumber & &
Tr_{q(2)_{_{}}}(RXR^{-1}) \; = \; Tr_{q(2)}(R^{-1}XR) \; = \;
I \cdot Tr_q X \; ,
\\
\lb{tr-add}  & &
Tr_{q(1,2)_{_{}}} (R f(X,R) R^{-1^{^{}}}) \; = \; Tr_{q(1,2)} \, f(X,R) \; ,
\\
\nonumber & &
Tr_{q(2)} R^{\pm 1^{^{}}} \; = \; q^{\pm N} I  \;\;\; ,
\;\;\; Tr_{q} I \; = \; [N]_q \; .
\ea
Here the index in parantheses denotes the number of the matrix space
in which the operation $Tr_q$ acts, and
$[N]_q = { q^N - q^{-N} \over \lambda}$.  \medskip

{\bf B.} {\it The ordering condition.} We suppose that
multiplication in the differential algebra is defined
by relations quadratic in $T$, $\Omega$ and $L$ and these
relations allow us to order lexicographycally any quadratic
monomial of the generators. Moreover, they should allow
the {\it unique} ordering for any higher order monomial of
$T$, $\Omega$ and $L$. The latter is the so-called
{\it diamond (or confluence) condition} ( see e.g. \cite{Bergman}).
It guarantees us that the Poincar\'e series of the classical
differential algebra do not change under quantization. The direct
check of this condition consists in the use of the
Diamond Lemma \cite{Bergman}. Such calculations appear to
be very cumbersome
already in the $N=2$ case
(see discussion in subsection 3.8 of the Ref. \cite{Man2}).
and it seems hard to generalize them for any $N$.
The alternative way we shall advocate here is in noticing that
the quadratical relations for $T$, $\Omega$, $L$ express in fact
the action of
some representation of the braid group on the differential algebra.
The diamond condition is then the consequence of the braid
group defining relations and, hence, it should follow from
the general properties (\ref{YB}), (\ref{H}) of the $R$-matrix.
Examples of such formal $R$-matrix manipulations are presented
in \cite{IP2} and in Section 2 of the present paper. \medskip

{\bf C.} The last but not the least condition is
that the differential algebra is to be supplied with the
differential complex sructure. In other words, we should
define the ${\bf C}$-linear differential mapping $d$ on it.
Taking into account the discussion above we choose the
following set of its characteristic properties:
\begin{itemize}
\item $d$ is of degree $1$ with respect to the natural ${\bf Z}$-grading
on the algebra of the differential forms;
\item $d$ satisfies the {\it nilpotence condition}: $d^2 = 0$.
\end{itemize}

Now let us proceed to the construction of such differential algebra.

\section{The differential algebra}
\setcounter{equation}0

We arrange the main result of this section in
\medskip \\
{\bf Theorem 1:} \it
For general values  of  the
deformation  parameter $q$
($[2]_q\neq0$, $[N]_q\neq~0$,
$[N]_q \neq - \lambda q^N$,
$[N\pm~1]_q\neq\pm~q^{N\mp~4}$)
the $GL_q(N)$-differential algebra  defined as
\ba
\lb{tt}
R\, T\, T' &=& T\, T'\, R  \; ,
\\
\lb{oo}
R\, \Omega \, R\, \Omega \; + \; \Omega \, R \, \Omega \, R^{-1}  &=&
\kappa_q ( \Omega^2 \; + \; R \, \Omega^2 \, R ) \; ,
\\
\lb{to}
R\, \Omega \, R^{-1} T &=& T\, \Omega' \; ,
\\
\lb{ll}
R\, L \, R \, L &=& L \, R \, L \, R \; ,
\\
\lb{tl}
R \, L \, R \, T &=& q^{{2 \over N}} T \, L' \; ,
\\
\lb{ol}
R^{-1} \Omega \, R \, L &=& L \, R \, \Omega \, R^{-1} \; ,
\ea
where
\be
\lb{kappa}
\kappa_q = {\lambda q^N \over [N]_q + \lambda q^N} \; ,
\ee
admits the consistent reduction to $SL_q(N)$.
This reduction is achieved by adding three more relations
\be
\lb{quot}
det_q\, T = 1 \; , \;\;\;\; Tr_q\, \Omega = 0 \; , \;\;\;\;
Det\, L = 1 \; ,
\ee
to (\ref{tt})-(\ref{ol}). Here
\be
\lb{Det}
Det\, L = q^{1-N} (R_1 R_2 \ldots R_{N-1} L_1)^N \epsilon_{q}^{1 \ldots N}
= q^{1-N} (L_1 R_1 R_2 \ldots R_{N-1})^N \epsilon_{q}^{1 \ldots N} \; .
\ee
\rm
\medskip
{\bf Proof:} It is not difficult  to check the bicovariance condition
for (\ref{tt}-\ref{quot}) by using the commutation properties of
$T$s (\ref{RTT}) and the definitions for left and right
transitions on the quantum group (\ref{comult}),(\ref{ad}).
Here we only mention the transformation properties of $Det\,L$:
$$
\delta_L(Det\,L) \; = \; 1 \otimes Det\, L \; , \;\;\;\;
\delta_R(Det\,L) \; = \; Det\, L \otimes 1 \; .
$$

The validity of the ordering condition for the quadratic monomials
of $T$, $\Omega$, $L$ is to be verified by rewriting relations
(\ref{tt}-\ref{ol}) in matrix components. We can rather convince
ourselves in this by noticing that relations (\ref{tt}), (\ref{oo}),
(\ref{ll}) contain the right number of the commutation relations
for $T$s, $\Omega$s and $L$s due to their symmetry property
$$
\begin{array}{l}
P_{q}^{\pm} ( RTT' - TT'R ) P_{q}^{\pm} \; \equiv \;
P_{q}^{\pm} ( RLRL - LRLR ) P_{q}^{\pm} \; \equiv \; 0 \; ,
\\
P_{q}^{\pm} \left( R\Omega R\Omega + \Omega R\Omega R^{-1} -
\kappa_q(\Omega^2 + R\Omega^2 R) \right) P_{q}^{\mp} \; \equiv \; 0 \; .
\end{array}
$$
Here $P_{q}^{\pm} \, = ( \pm R + q^{\mp 1} )/[2]_q$ are
the quantum symmetrizer and antisymmetrizer, respectively
(see \cite{Jimbo,FRT}).

Now, let us concentrate on checking the diamond condition
for monomials cubic in $T$, $\Omega$ and $L$. First, we
choose the suitable full set of such monomials:
\be
\lb{set}
\begin{array}{ccccc}
(R'\,R\,\Omega)^3\, , & T(R'\,\Omega)^2\, , & R\,T\,T'\,\Omega'' \, , &
R'\,R^{-1}\Omega\,R'^{-1}R\,\Omega\,R'\,R\,L\, , & T\,T'\,T'' \,
\\
(R'\,R\,L)^3\, , &  T(R'\,L)^2\, , & R\,T\,T'\,L''\, , &
R'^{-1}R^{-1}\Omega(R'\,R\,L)^2\, , & T\,\Omega'\,R'\,L\,R' \, .
\end{array}
\ee
Here $T'' = T_3 = I^{\otimes 2} \otimes T$ and the same is for
$\Omega''$ and $L''$.
The combinations (\ref{set}) are constructed so
that one can apply the 'commutation rules` (\ref{tt}-\ref{ol})
for any adjacent pair of the generators entering into them.
We interpret this operation as the ($q$-)permutation of a
pair of generators. Applying the $q$-permutations three times to
the monomials (\ref{set}) we arrange their entries in an inverse
order. Obviously, this reodering can be performed in two
different ways, depending on whether we first permute
the left pair of generators or the right one.
The diamond condition states that in both cases the result
must be the same. We demonsrate how the calculations proceed
in the most complex case of the $(R'\,R\,\Omega)^3$-reodering.
This example was already considered in \cite{IP2} and here we
present a simpler derivation.

The calculations proceed as follows:
\be
\lb{121}
\begin{array}{lll}
(R'\,R\,\Omega)^3 \; = & R\,R'
\underline{R\,\Omega\,R\,\Omega}
R'\,R\,\Omega & \\
& \quad\quad\downarrow^{1\leftrightarrow 2 \;perm.}& \\
&- R\,\Omega\,R\,R'
\underline{R\,\Omega\,R\,\Omega}
R'^{-1}
& + \; \kappa_qR\,R'\,(\Omega^2+R\,\Omega^2\,R)\,R'\,R\,\Omega \\
& \quad\quad\downarrow^{2\leftrightarrow 3 \;perm.}& \\
& \underline{R\,\Omega\,R\,\Omega}
R'\,R\,\Omega\,R^{-1}R'^{-1}
& - \; \kappa_qR\,\Omega\,R\,R'(\Omega^2 + R\,\Omega^2\,R)R'^{-1} \\
& \quad\quad\downarrow^{1\leftrightarrow 2 \;perm.}& \\
& - \Omega\,R\,R'\,\Omega\,R\,R'^{-1}\Omega\,R^{-1}R'^{-1}
& + \; \kappa_q(\Omega^2  +  R\,\Omega^2\,R)R'\,R\,\Omega\,R^{-1}R'^{-1}
\; ,
\end{array}
\ee
and in another way
\be
\lb{212}
\begin{array}{lll}
(R'\,R\,\Omega)^3 \; = & R'\,R\,\Omega\,R\,R'
\underline{R\,\Omega\,R\,\Omega} & \\
& \quad\quad\downarrow^{2\leftrightarrow 3 \;perm.}& \\
&- R'
\underline{R\,\Omega\,R\,\Omega}
R'\,R\,\Omega\,R^{-1}
& + \; \kappa_qR'\,R\,\Omega\,R\,R'(\Omega^2 + R\,\Omega^2\,R) \\
& \quad\quad\downarrow^{1\leftrightarrow 2 \;perm.}& \\
& \Omega\,R\,R'
\underline{R\,\Omega\,R\,\Omega}
R'^{-1}R^{-1}
& - \; \kappa_qR'(\Omega^2 + R\,\Omega^2\,R)R'\,R\,\Omega\,R^{-1} \\
& \quad\quad\downarrow^{2\leftrightarrow 3 \;perm.}& \\
& - \Omega\,R\,R'\,\Omega\,R\,R'^{-1}\Omega\,R^{-1}R'^{-1}
& +\; \kappa_q\Omega\,R\,R'(\Omega^2 + R\,\Omega^2\,R)R'^{-1}R^{-1}
\; .
\end{array}
\ee
Here we employ subsequently Eqs. (\ref{YB})
and (\ref{oo}) through all calculations. It remains to compare
the $\kappa_q$-terms arising under transformations (\ref{121}) and
(\ref{212}). Here we need one more formula \cite{IP2}
\be
\lb{o2o}
R\,\Omega^2 R\,\Omega \; - \; \Omega\,R\,\Omega^2 R \; = \; 0\; .
\ee
It is derived as follows: denoting the l.h.s. of (\ref{o2o})
as $U$ and using (\ref{oo}) twice we get
\be
\lb{U}
U \; +\;  \kappa_q R U R  \; =\; 0\; .
\ee
Now, dividing $U$ into a sum of $q$-symmetric and $q$-antisymmetric
parts $U_{\pm}$:
$$
U_{\pm} = U \pm R U R^{\pm 1} \; , \;\;\;
P_{q}^{\pm} U_{+} P_{q}^{\mp} = P_{q}^{\pm} U_{-} P_{q}^{\pm} = 0 \; ,
\;\;\;
U = { ( 1 + R^{-2} ) \over [2]_{q}^{2}}
\left( U_{+} \, + \, U_{-} \right) \; ,
$$
we transform (\ref{U}) into a couple of relations
$$
( I + \kappa_q R^2 ) U_{+} = 0 \; , \;\;\;\;
( 1 - \kappa_q ) U_{-} = 0 \; .
$$
Then, under restrictions $(1 + \kappa_q R^2) \not\sim P_{q}^{\pm}$,
$\kappa_q \neq 1$, or, equivalently, $[N\pm 1]_q \neq \pm q^{N\mp 4}$,
$[N]_q \neq 0$ we get the desired relation (\ref{o2o}).

Now, one can compare the $\kappa_q$-terms in (\ref{121}) and
(\ref{212}), turning all the $\Omega^2$ entries to the left.
The result is the same in both cases and, thus, the diamond
condition on $(R'\,R\,\Omega)^3$ is satisfied. The same calculations,
although simpler, can be carried out for all other monomials
of (\ref{set}), and we leave them as an exercise.

It remains to check the consistency of the $SL_q(N)$-reduction.
The centrality of $det_q\,T$ is easily proved  by the use
of relation (\ref{eR}). Next, the application of $Tr_{q(2)}$
to Eq. (\ref{oo}) and subsequent use of the Hecke relation (\ref{H})
give
$$
\left[ Tr_q\Omega\, ,\, \Omega \right]_{+} \; + \;
\lambda q^N \Omega^2 \; = \; \kappa_q \left( [N]_q + \lambda q^N \right)
\Omega^2 + \kappa_q Tr_q\Omega^2 \;,
$$
whereof we conclude that $Tr_q \Omega$ anticommutes with $\Omega$
under conditions that \\
-- the parameter $\kappa_q$ is chosen as in (\ref{kappa}); \\
-- the quadratic scalar combination $Tr_q \Omega^2$ identically
vanishes. \\
The last is the direct consequence of (\ref{oo}). It is derived
like follows. Applying $Tr_{q(1,2)} (\ldots)$ and
$Tr_{q(1,2)} (\ldots R^{-1})$ operations  to (\ref{oo}) and
using (\ref{tr-add}),(\ref{H}) we get the system of linear relations
on the quadratic scalars $(Tr_q \Omega)^2$
and $Tr_q \Omega^2$:
$$
\begin{array}{c}
2(Tr_q \Omega)^2 \, - \, [N]_q \kappa_q Tr_q \Omega^2 \; = \; 0 \; , \\
- \lambda (Tr_q \Omega)^2  \, + \, \left( 2q^N - \kappa_q
(q^N + q^{-N}) \right) Tr_q \Omega^2 \; = \; 0 \; .
\end{array}
$$
The determinant of this system:
${q^N [2]_{q}^{2} [N]_q \over [N]_q + \lambda q^N }$ -- does not vanish
under conditions of the Theorem and, hence, we conclude
$$
(Tr_q \Omega)^2 \, = \, Tr_q \Omega^2 \, = \, 0\; .
$$
Then,
applying the $Tr_{q(2)}$ operation to (\ref{to}), (\ref{ol})
we find that $Tr_q \Omega$ is the (graded) central element
in the algebra (\ref{tt})-(\ref{kappa}).

Finally, to construct the central element from $L$s, we use
the following trick suggested in \cite{AF2,SWZ,DJSWZ} (see also
\cite{Zum2}). Consider the matrix $Z = LT$. It behaves like $T$
under the left and right transitions in $GL_q(N)$. Moreover,
it possesses the similar algebraic properties:
\ba
\nonumber
R\, Z\, Z'  &=& Z\, Z'\, R \; , \\
\nonumber
R^{-1} \Omega\, R\, Z &=& Z\, \Omega' \; , \\
\nonumber
R\, L\, R\, Z &=& q^{{2 \over N}} Z\, L' \; .
\ea
Hence, $Det L = det_q\,Z \cdot (det_q\,T)^{-1}$ is central in the algebra
(\ref{tt})-(\ref{ol}). Now, let us show that $Det L$ indeed
depends only on $L$:
$$
\begin{array}{ll}
\epsilon_{q}^{1 \ldots N} \cdot Det L &= (L_1T_1)(L_2T_2)
\ldots (L_N T_N)
\epsilon_{q}^{1 \ldots N} \cdot (det_q\,T)^{-1}
\\
&= q^{N-1} L_1(R_1L_1R_1) \ldots (R_N \ldots R_1 L_1
R_1 \ldots R_N) \epsilon_{q}^{1 \ldots N}  \;.
\end{array}
$$
The expression (\ref{Det}) for $Det L$ is then extracted by using
(\ref{ll}), (\ref{YB}) and performing induction in $N$. {\bf Q.E.D.}

{\bf Comment:} Among relations (\ref{tt})-(\ref{ol}) only (\ref{to})
is a completely new relation. Formula (\ref{oo}) were proposed
for $N=2$ case in \cite{IP1} and for general $N$ in \cite{IP2}
as commutation relations for Cartan $1$-forms on $SL_q(N)$.
Formulae (\ref{ll}), (\ref{tl}) appeared in \cite{AF}
as the algebra of functions on the cotangent bundle of
$GL_q(N)$. The algebra of vector fields (\ref{ll})-(\ref{ol})
is suggested in \cite{SWZ,Zum2} for the differential calculus
on $GL_q(N)$ and $SL_q(N)$. The definition of quantum determinant
$Det L$ can also be found in these works and in \cite{DJSWZ}.
Note also the recent work \cite{AAM} where the external algebra
(\ref{tt})-(\ref{to}) has been given in components for $N=2$ case.
A really new point in our approach is that all these formulae
are consistently combined into a single algebra.

{\bf Remark 1:} Besides the algebra (\ref{tt})-(\ref{quot})
there exist three more differential algebras on $SL_q(N)$.
They can be obtained from (\ref{tt})-(\ref{ol}) by the
substitutions of two types:
\ba
\lb{s1}
{\bf S1\,:}&& \qquad R \leftrightarrow R^{-1}\; , \;\;\kappa_q
\leftrightarrow \kappa_{{1 \over q}} \quad {\rm in \;\; (\ref{oo})} \; ; \\
\lb{s2}
{\bf S2\,:}&& \qquad R \leftrightarrow R^{-1} \; , \;\;
q \leftrightarrow q^{-1} \quad\;\; {\rm in \;\; (\ref{to})-(\ref{ol})
, \; (\ref{Det}).} \; .
\ea
For $N=2$ the substitution (\ref{s1}) is trivialised. Indeed, the relations
(\ref{oo}) and
\bf S1$\cdot$\rm(\ref{oo}) in case $N=2$  differ by the term
proportional to $P_{q}^{-}(\Omega^2 + R\Omega^2 R)\, \sim \,
P_{q}^{-}\Omega^2
P_{q}^{-}\, \sim \,
Tr_q \Omega^2 \cdot P_{q}^{-}$
and since the scalar relation $Tr_q\Omega^2$ is contained both in (\ref{oo})
and
\bf S1$\cdot$\rm(\ref{oo}), it follows that relations (\ref{oo}) and
\bf S1$\cdot$\rm(\ref{oo}) for $N=2$ are identical. This result agrees
with the statement of \cite{AAM} that there exist only two
different external algebra structures on $SL_q(2)$.
We should stress here that this mechanism does not
work for $N>2$, where we have $4$ noncoinciding
differential algebras.

{\bf Remark 2\,:} The very limited number of the $q$-deformations
for the differential calculus on $SL(N)$ seems to be
an effect of the simplicity property of $SL(N)$.
By contrast, one could derive a lot of the quantized
versions for $GL(N)$ case. For instance,
if we take away the condition of the existence of
$SL_q(N)$-reduction, then it is no need of fixing
parameters $\kappa_q$ and $q^{{2 \over N}}$ in Eqs. (\ref{oo}),
(\ref{tl}). Another possibility is to use for $T$ and $\Omega$ the
commutation rules different from
(\ref{to}):
\be
\lb{another}
R\,\Omega\,R\,T \; = \; T\,\Omega' \; .
\ee
Algebras of that type were considered in
\cite{Man2,Malt,Sud,Schir,Mul,SWZ,Tzy,Zum2,IP2}.

{\bf Remark 3\,:} A few words on the interpretation of
the basic vector fields $L$ are in order. It is very natural
to suppose that the algebra of the classical vector fields $V$
behaves under the quantization like $U_q g$ and, hence, is not
quadratic. On the other hand, simple quadratic relations
are achieved for another-type generators $L^+$, $L^-$ \cite{FRT}
and $L$ \cite{RS,AFS,AF2,KS}. These generators constitute finite shifts
on the quantum group and can be viewed as some `exponentiated'
form of infinitesimal vector fields $L = I + \lambda V + O(\lambda^2)$.
That is why the $SL_q(N)$ reduction for $L$ is performed
not by the $Tr_q$-like condition but by its exponentiated $Det$-like form.
It is also natural from this point of view
that the quantities $Z = LT$ obtained from $T$s by finite
$L$-shifts behave algebraically like $T$s.

\section{Exterior derivative}
\setcounter{equation}0

We shall define the differential mapping $d$ on external algebra
(\ref{tt})-(\ref{to}) in a constructive way.
\begin{enumerate}
\item
Define the action of $d$ on the generators $T$ and $\Omega$
as
\be
\lb{dt}
dT = \Omega\,T \; , \;\;\;\; d\Omega = \Omega^2 \; .
\ee
\item
For the Cartan $1$-forms we postulate the ordinary Leibnitz rule
to be satisfied
\be
\lb{oleib}
d \cdot \Omega = \Omega^2 - \Omega \cdot d
\ee
Using (\ref{o2o}) it is strightforward to check that
this prescription agrees with the commutation relations for $\Omega$s
(\ref{oo}). Becides, due to (\ref{oleib}) the action of exterior
derivative on $T$ and on any function $F$ of $\Omega$ is
nilpotent: $d^2T = d^2 F(\Omega) = 0$. Leaving
apart the mathematical reasonings, we would like to
stress that it is rather natural to keep the classical
Leibnitz picture for infinitesimal objects like $\Omega$.
\item
Using {\bf 1.} and {\bf 2.} we are able to calculate the
exterior derivative action on any monomial of $T$ and $\Omega$
which is of first order in $T$. Namely, we should
at first turn all $\Omega$s to the left by using commutation
relations (\ref{to}), and then apply (\ref{dt}) and (\ref{oleib}) to get:
$d(F(\Omega)T) = dF(\Omega)\,T + F(-\Omega)\Omega T$.
In this way we get automatically the consistency of
the differential mapping with the algebraic relations (\ref{to})
and the nilpotence of $d$ on any monomial of that type.
\item
The next step is to construct the differential mapping
for the general quadratic monomial of $T$: $TT'$.
We stress here that since under quantization we obtain
the finite shifts $L$ acting on $T$ rather than
differentiation, it is very reasonable to get the
modified Leibnitz rules for $T$. The action of $d$
should take into response the algebraic relations (\ref{tt}):
$$
R\,d(T\,T') \; = \; d(T\,T')\,R \; .
$$
Note also that the expression for $d(T\,T')$ must be of $1$-st order
in $\Omega$. The general ansatz satisfying both these conditions
reads
\be
\lb{ans2}
d(T\,T') = f(R)(\Omega + R\,\Omega\,R) T\,T' \; .
\ee
Here $f(R)$ is a function of $R$ and the combination
$\Omega + R\,\Omega\,R$ commutes with the $R$-matrix
due to Hecke conditions (\ref{H}). The exact form of the
function $f(R)$ is dictated by the nilpotence condition:
$$
0 \; =\; d^2(T\,T') \; = \; f\left\{ (\Omega^2 + R\,\Omega^2 R) \, - \,
f(\Omega + R\,\Omega\,R)^2 \right\}T\,T'\; .
$$
Using (\ref{oo}) it is straightforward to abtain
$$
(\Omega + R\,\Omega\,R)^2 \; = \; (I + \kappa_q R^2)(\Omega^2
+ R\,\Omega^2 R) \; ,
$$
and, hence, $d$ is nilpotent on $T\,T'$ if we put\footnote{
Note that under restrictions of Theorem $1$ the matrix $(I + \kappa_q R^2)$
is invertible.}
\be
\lb{f}
f(R) = (I + \kappa_q R^2 )^{-1} \; .
\ee
Using (\ref{to}), (\ref{dt}), (\ref{oleib}), (\ref{ans2})
we can obtain now how $d$ acts on any monomial of $T$ and $\Omega$
that is
quadratic in $T$, and again
the nilpotence of $d$ is guaranteed by  (\ref{oleib}).
\end{enumerate}

Thus, we gave the detailed consideration
of the  the first few steps in constructing
the differential mapping $d$.
Generalizing this procedure to  monomials of
any order in $T$ we get \medskip \\
{\bf Theorem 2\,:} \it
For the external algebra (\ref{tt})-(\ref{to})
presented in Theorem $1$ there exists the left acting
differential mapping $d$, defined by (\ref{dt}), (\ref{oleib})
and
\be
\lb{dtk}
d(T_1 T_2 \ldots T_k) = \left\{ I + \kappa_q(S_k(I) - I) \right\}^{-1}
S_k(\Omega) T_1 T_2 \ldots T_k \; ,
\ee
where
\be
\lb{s}
S_k(X) = X \; + \;
\sum_{i=1}^{k-1} R_i \ldots R_2 R_1 X R_1 R_2 \ldots R_i \; .
\ee
In particular,
\ba
\lb{ddetT}
d(det_q\, T) &=& {1 \over q^{N-1} (1 - \kappa_q) \, + \, [N]_q \kappa_q}
Tr_q\, \Omega \; det_q\, T\; ,
\\
\lb{dtro}
d(Tr_q\, \Omega) &=&  Tr_q\, \Omega^2 \; = \; 0 \; ,
\ea
which guarantees the compatibility of $d$ with the reduction
conditions (\ref{quot}).
 This differential mapping is commutative with the action
of the basic vestor fields $L$:
\be
\lb{dl}
\left[ d\, , \, L \right] = 0 \; .
\ee
\rm
{\bf Proof:}
As in the case $k=2$ we start with the following general ansatz
\be
\lb{ans-k}
d(T_1 T_2 \ldots T_k) = f_k\,S_k(\Omega) T_1 T_2 \ldots T_k \; .
\ee
Here $f_k$ is a function of $R_1, \ldots , R_k$ to be specified below,
and
\be
\lb{symm}
R_i f_k = f_k R_i \; , \;\;\;\; R_i S_k(\Omega) = S_k(\Omega) R_i \; ,
\;\;\;\; i=1, \ldots , k-1 \;.
\ee
The first of relations (\ref{symm}) is the restriction on the
possible form of  $f_k$, while the last is the direct consequence
of the Yang-Baxter equation (\ref{YB}) and the Hecke condition (\ref{H}).
In virtue of Eqs. (\ref{symm}) we have
$$
R_i\, d(T_1 \ldots T_k) = d(T_1 \ldots T_k) \, R_i \; , \;\;\;\;
i = 1, \ldots , k-1 \; ,
$$
and, thus, ansatz (\ref{ans-k}) is compatible with relations (\ref{tt})
of the external algebra. The nilpotence condition $d^2(T_1 \ldots T_k) = 0$
leads to the relation
$$
S_k(\Omega^2) - f_k ( S_k(\Omega))^2 = 0 \; .
$$
It remains to compute the quantity $(S_k(\Omega))^2$.
This calculation is based on the essential use of Eqs. (\ref{oo}),
(\ref{YB}) and (\ref{H}) is rather lengthy. We present here
the result
$$
(S_k(\Omega))^2 = \left\{ I + \kappa_q(S_k(I) - I ) \right\} S_k(\Omega^2) \; .
$$
Hence, the function $f_k$ is to be chosen as in (\ref{dtk}).
Note that with this choice $f_k$ satisfies the conditions (\ref{symm}).

In obtaining the formula (\ref{ddetT}) one should employ the properties
(\ref{e}) of the $q$-deformed Levi-Civita tensor, and also
$$
\epsilon_{q}^{1\ldots N} S_N(X) = q^{1-N} Tr_q\, X\,
\epsilon_{q}^{1\ldots N} \; .
$$
The check of the compatibility of condition (\ref{dl}) with the algebra
(\ref{ll})-(\ref{ol}) is straightforward. {\bf Q.E.D.} \medskip \\
{\bf Remark 1:} Using (\ref{dt}), (\ref{oleib}), (\ref{dtk}),
and (\ref{to}), (\ref{oo}) one can derive
the explicit form of the modified Leibnitz rules.
These rules appear in the modified form for $T$, $\Omega$-polynomials,
for which the left acting exterior derivative should
cross $T$ under evaluation. For the quadratic polynomials
we have
\ba
\nonumber
d(T\,T') &=& (I + \kappa_q R^2)^{-1} \{ R^2 dT\, T'\, + \, T\, dT' \} \\
\nonumber
d(T\, \Omega') &=& (1 - \kappa_q) T\, d\Omega' \;+\; dT\,\Omega'
\;+\; \{ (1-\kappa_q)R^2 - I \}\Omega^2\,T\;.
\ea
Here the term $\Omega^2\,T$ may be treated either as $d\Omega\,T$
or as $\Omega\,dT$. Note that the operator $R^2$, being the generating
element  of the braid group $B_2$, plays a particular role
in these formulae. This observation is further approved
if we evaluate the action of $d$ on the monomials of any
order in $T$:
\be
\lb{bdt}
d(T_1 \ldots T_k) = \left\{ I + \kappa_q \sum_{i=1}^{k-1}
B_{k,i} \right\}^{-1} \, \sum_{i=1}^{k} B_{k,i} \, T_1 \ldots dT_i
\ldots T_k \; ,
\ee
$$
B_{k,i} = (R_i R_{i+1} \ldots R_{k-1})(R_{k-1} R_{k-2} \ldots R_i ) \; ,
\;\;\; B_{k,k} = I^{\otimes k} \; , \;\;\; i=1, \ldots , k-1 \; .
$$
Here $\{B_{k,i} \}_{i=1}^{\;\;\;k}$ is the set of generating elements for
the braid group $B_k$. \medskip \\
{\bf Remark 2:} Note that in constructing the differential mapping
$d$, essential are
the self-commutation relations for $T$ (\ref{tt})
and $\Omega$  (\ref{oo}). The explicit form of the cross-commutation
relations for $T$ and $\Omega$
(\ref{to}) is not relevant. We only should be aware of
that these relations allow us to turn all $\Omega$s to the left
in any monomial of $T$ and $\Omega$. Thus, the algorithm described can be
applied equally to the external algebras considered in
\cite{Man2,Malt,Sud,Schir,Mul,SWZ,Tzy,Zum2,IP2} and satisfying the
cross-multiplication relations of the type (\ref{another}). In this way
one can search for all the external algebraic structures
on $GL_q(N)$ compatible with the ordinary Leibnitz prescriptions.
It turns out that only two external algebras obtained in the references
above satisfy this condition. The first of these algebras
is defined by relations (\ref{tt}), (\ref{another}) and (\ref{oo})
where one must put $\kappa_q = 0$. The second algebra is
obtained from the first  if one makes substitution $R \leftrightarrow
R^{-1}$ in all formulae. This result agrees with the
quasiclassical considerations of Ref. \cite{AM}.

\section*{Acknowledegments}
We would like to thank
A.P.Isaev for fruitful collaboration and stimulating discussions.


\begin{thebibliography}{99}

\bibitem{AF} Alekseev, A.Yu., Faddeev, L.D.:
Commun. Math. Phys. {\bf 141} (1991) 413-422.

\bibitem{AF2} Alekseev, A.Yu., Faddeev, L.D.:
Involution and dynamics in the system $q$-deformed top.
Zapiski Nauch. Sem. LOMI {\bf 200} (1992) 3.

\bibitem{AFS} Alekseev, A.Yu., Faddeev, L.D., Semenov-Tian-Shansky, M.A.:
Commun. Math. Phys. {\bf 149} (1992) 335-345.

\bibitem{AAM} Aref'eva, I.Ya., Arutyunov, G.E., Medvedev, P.B.:
Poisson-Lie structures on the external algebra of $SL(2)$
and their quantization. Preprint SMI (1994) and hep-th/9401127.

\bibitem{AM} Arutyunov, G.E., Medvedev, P.B.:
Quantization of external algebra on a Poisson-Lie group.
Preprint SMI-11-93 and hep-th/9311096.

\bibitem{AC} Aschieri, P., Castellani, L.:
Intern. Journ. Mod. Phys. A {\bf 8} (1993) 1667-1706;
Castellani, L., R-Monteiro, M.A.: Phys. Lett. {\bf 314}B (1993) 25-30.

\bibitem{Bergman} Bergman, G.M.:
Adv. in Math. {\bf 29} (1978) 178-218.

%\bibitem{Bernard} Bernard, D.: Progr. Theor. Phys. Suppl. {\bf 102}
%(1990) 49-66.

\bibitem{CSWW} Carow-Watamura, U., Schlieker, M., Watamura, S.,
Weich, W.: Commun. Math. Phys. {\bf 142} (1991) 605.

\bibitem{DJSWZ} Drabant, B., Jur\v{c}o, B., Schlieker, M.,
Weich, W., Zumino, B.:
Lett. Math. Phys. {\bf 26} (1992) 91-96.

\bibitem{FRT} Faddeev, L.D., Reshetikhin, N.Yu., Takhtajan, L.A.:
Quantization of Lie groups and Lie algebras (in Russian).
Algebra i Analiz
{\bf 1} (1989) 178-206.

\bibitem{Faddeev} Faddeev, L.D.: Lectures on Int. Workshop
`Interplay between Mathematics and Physics'. Vienna 1992 (unpublished)

\bibitem{IM} Isaev, A.P., Popowicz, Z.:
Phys.Lett. {\bf 281}B (1992) 271; \\
Isaev, A.P., Malik, R.P.: Phys.Lett. {\bf 280}B (1992) 219.

\bibitem{IP1}  Isaev, A.P., Pyatov, P.N.:
Phys.Lett. {\bf 179}A (1993) 81-90.

\bibitem{IP2} Isaev, A.P., Pyatov, P.N.: Covariant differential
complexes on quantum linear groups. Preprint JINR E2-93-416
and hep-th/9311112.

\bibitem{Jimbo}
Jimbo, M.:  Lett. Math. Phys. {\bf 11} (1986) 247-252.

\bibitem{Jurco} Jur\v{c}o, B.:
Lett. Math. Phys. {\bf 22,} 177 (1991) 177-186.

\bibitem{KS} Kulish, P.P., Sasaki, R.: Progr. of Theor. Phys.
{\bf 89} (1993) 741-761.

\bibitem{Malt} Maltsiniotis, G.: C.R.Acad.Sci. Paris {\bf 331}
(1990) 831; \\
Calcul diffe'rentiel sur le groupe line'arie quantique.
Preprint ENS (1990).

\bibitem{Man0} Manin, Yu.I: Quantum Groups and Noncommutative Geometry.
Montreal University Preprint CRM-1561 (1989).

\bibitem{Man1} Manin, Yu.I.:
Commun. Math. Phys. {\bf 123} (1989) 163.

\bibitem{Man2} Manin, Yu.I.:
Teor. Mat. Fiz. {\bf 92} (1992) 425-450.

\bibitem{Mul} M\"{u}ller-Hoissen, F.: Journ. Phys. A {\bf 25} (1992) 1703;
M\"{u}ller-Hoissen, F., Reuten, C.: ibid.
{\bf 26} (1993) 2955.

\bibitem{PW} Podle\'s, P., Woronowicz, S.L.: Commun. Math. Phys.
{\bf 130} (1990) 381-431.

\bibitem{Reshet1} Reshetikhin, N.Yu: Algebra i Analiz {\bf 1}
169 (1989)

%\bibitem{Reshet2} Reshetikhin, N.Yu.: Leningrad Math. J. {\bf 1}
%(1990) 491.

\bibitem{RS} Reshetikhin, N.Yu., Semenov-Tian-Shansky, M.A.:
Lett. Math. Phys. {\bf 19} (1990) 133.

\bibitem{Schir} Schirrmacher, A. In: Gielerak, R., Lukierski, J.,
Popowicz, Z. (eds.)
Groups and Related Topics. Proceedings, Wroclaw 1991, p. 55.
Kluwer Academic Publishers 1992.

\bibitem{SVZ} Schmidke, W.M., Vokos, S.P., Zumino, B.:
Z. Phys. C {\bf 48} (1990) 249-255.

\bibitem{SWZ} Schupp, P., Watts, P., Zumino, B.:
Lett. Math. Phys. {\bf 25} (1992) 139-147;
Commun. Math. Phys. {\bf 157} (1993) 305.

\bibitem{Sud} Sudbery, A.:  Phys. Lett. {\bf 284}B (1992) 61;
Math. Proc. Camb. Phil. Soc. {\bf 114} (1993) 111.

\bibitem{Tzy} Tzygan, B.:
Notes on differential forms on quantum groups.
Penn. Univ. Preprint, 1992.

\bibitem{WesZum} Wess, J., Zumino, B.: Nucl. Phys. B (Proc. Suppl.)
{\bf 18} (1990) 302.

\bibitem{Wor1} Woronowicz, S.L.: Publ. RIMS, Kyoto University {\bf 23}
(1987) 117-181.

\bibitem{Wor2} Woronowicz, S.L.:
Commun. Math. Phys. {\bf 122}
(1989) 125-170.

\bibitem{Zum1} Zumino, B.: Introduction to the Differential
Geometry of Quantum Groups. Preprint University of California
UCB-PTH-62/91 (1991)
and in Proc. of X-th IAMP Conf., Leipzig 1991, p. 20.
Springer-Verlag 1992.

\bibitem{Zum2} Zumino, B.: Differential Calculus on quantum spaces
and Quantum Groups. Preprint LBL-33249 and UCB-PTH-92/41 (1992).

\end{thebibliography}
\end{document}